\documentclass[twocolumn,showpacs,superscriptaddress,amsmath,amssymb,floatfix,pra]{revtex4}

\usepackage{graphicx}
\usepackage{dcolumn}
\usepackage{bm}
\usepackage[usenames]{color}

\begin{document}

\title{Designing potentials by sculpturing wires}

Phys. Rev. A \textbf{75}, 063604 (2007)

\author{Leonardo Della Pietra}
\author{Simon Aigner}
 \affiliation{Physikalisches Institut, Universit\"{a}t Heidelberg, 69120 Heidelberg,Germany}
\author{Christoph vom Hagen}
 \affiliation{Physikalisches Institut, Universit\"{a}t Heidelberg, 69120 Heidelberg,Germany}
 \affiliation{Atominstitut der \"Osterreichischen Universit\"aten, TU-Wien, A-1020 Vienna, Austria}
\author{S\"{o}nke Groth}
 \affiliation{Physikalisches Institut, Universit\"{a}t Heidelberg, 69120 Heidelberg,Germany}
\author{Israel Bar-Joseph}
 \affiliation{Department of Condensed Matter Physics, The Weizmann Institute of science, Rehovot 7600, Israel }
\author{Henri J.Lezec}
 \affiliation{ISIS,Universit\'{e} Louis Pasteur, 8 all\'{e}e Gaspard Monge, 67083 Strasbourg, France }
\author{J\"org Schmiedmayer}
 \affiliation{Physikalisches Institut, Universit\"{a}t Heidelberg, 69120 Heidelberg,Germany}
 \affiliation{Atominstitut der \"Osterreichischen Universit\"aten, TU-Wien, A-1020 Vienna, Austria}

\begin{abstract}
Magnetic trapping potentials for atoms on atom chips are
determined by the current flow in the chip wires. By
modifying the shape of the conductor we can realize specialized
current flow patterns and therefore micro-design the trapping
potentials. We have demonstrated this by nano-machining an atom
chip using the focused ion beam technique. We built a trap, a barrier
and using a BEC as a probe we showed that by polishing the
conductor edge the potential roughness  on the
selected wire can be reduced. Furthermore we give different other
designs and discuss the creation of a 1D magnetic lattice on an
atom chip.
\end{abstract}


\maketitle

\section{Introduction}

Miniaturized atom optical elements on atom chips form the basis
for robust neutral atom manipulation~\cite{Fol02,Eur05}. The basic
design idea of microscopic magnetic traps and guides is to
superimpose the field of a current-carrying wire on the chip with
an external bias field. Atoms are trapped in the potential minimum
formed by the cancellation of the two fields. A straight wire and
a homogeneous bias field give a guide for atoms~\cite{Den99},
which can be transformed into a trap by changing the angle between
the current flow and the bias field. The standard solution is to
bend the wire. A U-shape results in a quadrupole, a Z-shape in a
Joffe trap~\cite{Fol02}.

In this paper we implement slight changes in the current path by
sculpturing the bulk of a lithographically patterned plane
conductor, as a mean to design the potential landscape. Even
though this results only in minute changes of the overall current
flow these are sufficient to alter significantly the local
potentials for ultra-cold atoms~\cite{PotSens}.

\section{Designing Potentials}

\begin{figure}
\begin{center}
\includegraphics[scale=0.43]{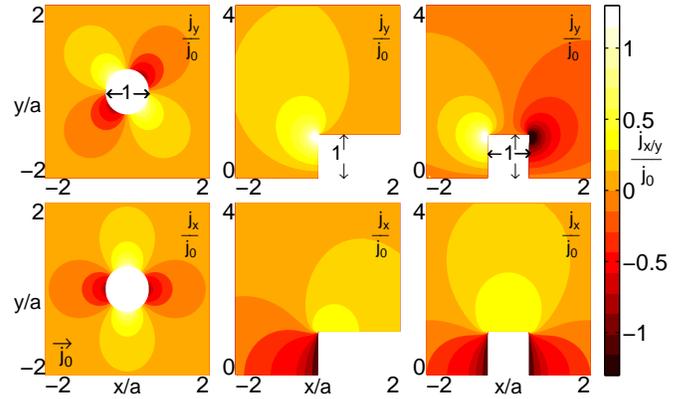}  \end{center}
   \caption{(Color online) \emph{Left}: a round cut in a conducting sheet creates current density perturbations
   parallel ($j_{x}$) and perpendicular ($j_{y}$) to the unperturbed flow ($j_0$). \emph{Center}: situation for
   a step at the edge of a semi infinite sheet. \emph{Right} a notch in the edge. In these
   last cases maxima and minima of $j_y$ have a singularity at the internal
   corners. Figures show a region of interest around defects of typical size $a$.
   \label{fig:wire}}
\end{figure}

We start by looking at holes and defects in a plane conductor
(fig.\ref{fig:wire}). The current flow around a circular hole in a
conducting sheet can be found analytically~\cite{Feyn}. For
general structures one needs numeric calculations. For a polygonal
cut in the vicinity of an edge of the conductor the
Schwarz-Christoffel transformation  can be used (conformal mapping
\cite{Conform}). The simplest is a single step in the edge: it
creates either a barrier or a trough, depending on the orientation
of the dominating longitudinal field component and on the
direction of the step.

The magnetic field of a general two dimensional current
distribution can be evaluated in the Fourier domain. Following
\cite{Roth}, the current densities' Fourier Transform components
in x and y direction  ($\hat{j}_{x/y}$) are defined by:
\begin{equation}
    \hat{j}_{x/y}(k_{x},k_{y})=\int_{-\infty}^{\infty}dx\,\int_{-\infty}^{\infty}dy\, j_{x/y}(x,y)e^{i(k_{x}x+k_{y}y)}
\label{eq:a}
\end{equation}
and the magnetic fields' Fourier transform:
\begin{eqnarray}
    \hat{B}_{x} & = & \;\;\,f(k_{x},k_{y},z,d)\,\hat{j}_{y}(k_{x},k_{y})\,\,\,\\
    \hat{B}_{y} & = & -f(k_{x},k_{y},z,d)\,\hat{j}_{x}(k_{x},k_{y})\nonumber
\end{eqnarray}
where the function $f$ acts as a low pass filter for $j_{x/y}$. For homogeneous conductors, with $j_{x/y}$ independent of $z$
\begin{equation}
    f(k_{x},k_{y},z,d)=\frac{\mu_{0}}{2}\cdot d\left(\frac{1-e^{-kd}}{kd}\right) e^{-kz}
\label{eq:c}
\end{equation}
where $k=\sqrt{k_x^2+k_y^2}$ is the absolute value of the wave vector, $z$ the distance to
the top of the current carrying surface and $d$ the thickness of
the layer. At the surface of a thin conductor ($kd\ll 1$)
$f=\mu_{0}d/2$.

Eq.~\ref{eq:c} shows that the magnetic field computation can be
effectively done by applying a low pass filter in Fourier space to
the current distribution. This nicely illustrates the limits for
potential design. In order to keep features up to a certain wave
vector $k_{max}=\frac{2\pi}{\lambda_{min}}$ and to maximize their
amplitude two requirements must be met:

\begin{itemize}
    \item As $f(k_{x},k_{y},z,d)\propto e^{-kz}$  the contributions of the
    current flow pattern are exponentially damped when receding from
    the wire. To achieve a modulation of a fraction $\eta_0$ of the
    maximum achievable field at a minimum structure size
    $\lambda_{min}$ one has to stay at distances
    $z<-\frac{\log{\eta_0}}{k_{\max}}$

    \item For finite thickness $d$, the contribution of current flowing
    deep below the surface of the sheet starts to drop off already
    inside the conductor.  Increasing $d$ leads to larger modulation
    but saturates in the limit of a thick conductor. To reach a
    fraction $\eta_1$ of the maximum achievable modulation, one has to
    satisfy $d>\frac{-\log{(1-\eta_1)}}{k_{max}}$
\end{itemize}

\section{Polishing wires}

\begin{figure}
\begin{center} \includegraphics[scale=0.43]{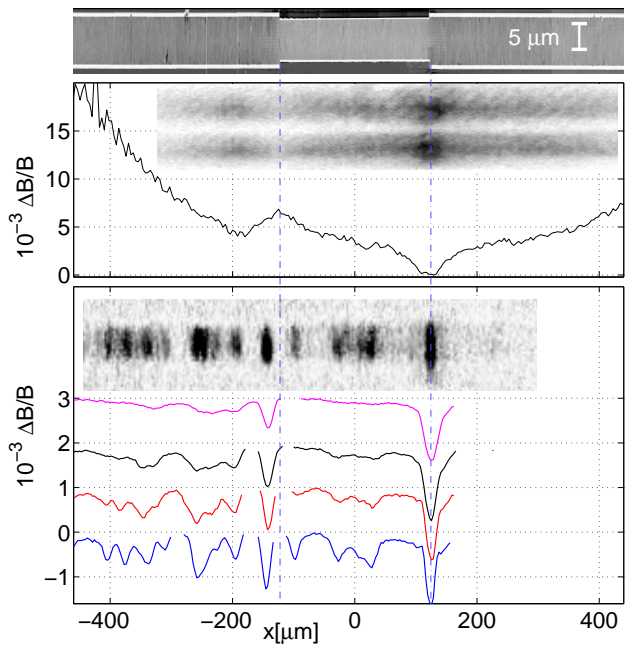} \end{center}
   \caption{(Color online) \emph{Top:} electron
   microscope picture of the polished wire. \emph{Center:}
   Overall form of the change ($\Delta B$) in the magnetic field minimum
   along the elongated trap in units of the total local wire field $B$
   (measured with thermal atoms at $z=35\,\mathrm{\mu m}$ from surface);
   insert: the thermal cloud (Time Of Flight: 2ms).
   \emph{Bottom:} measurement of the potential roughness using a 1D BEC
   at ${z=20,\,16,\,12,\,8\,\mathrm{\mu m}}$ (top to bottom).
   The curves are interrupted where roughness exceeds
   the chemical potential and shifted for visibility; insert: absorption image of a BEC
   around the wire's polished region;  $z=9\,\mathrm{\mu m}$, TOF=1.8ms.
   \label{fig:Polished-area}}
\end{figure}

Atom Chip wires can be easily sculptured with a focused ion beam
(FIB) technique, creating structures with high precision ($<20$
nanometers) and large aspect ratios (height/width$>30$)
\cite{FIB}. We experimentally demonstrated the power of this
technique by sculpturing a $10\,\mathrm{\mu m}$ wide and $2.5\,\mathrm{\mu m}$ thick
gold conductor (see fig.\ref{fig:Polished-area} top),
lithographically fabricated on a Si substrate by our standard
method~\cite{Gro04,fol00}. In our example we cut and polished the
vertical faces of this wire on both edges using the FIB over a
distance of $250\,\mathrm{\mu m}$.

This modification enables us to study two effects:

\begin{itemize}
\item At the limits of the polished regions we have steps in the
wire edge, as in the examples discussed above and in~\cite{DPie}.
These steps demonstrate that small deviations in the current flow
can be used to deliberately design potentials for tight traps and
barriers.
\item Comparing the potential roughness  in the polished
and in the untreated sections of the wire we get an indication of how
much of it stems from the~wire edges as proposed by~\cite{Wang04}
and how much from the irregular current flow in the wire itself as
suggested by~\cite{kru05}.
\end{itemize}

\begin{figure}[t]
\begin{center}
\includegraphics[width=8.6cm]{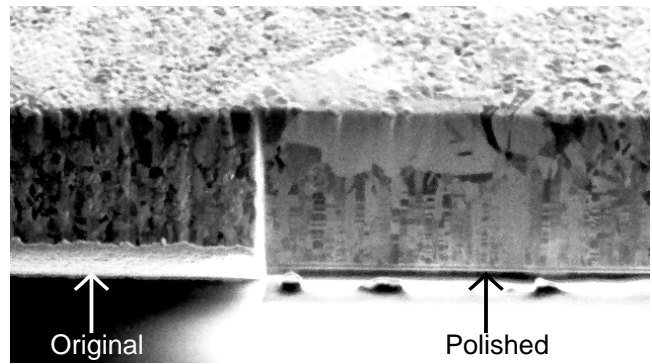}
\caption{\label{fig:micro}
A step in a wire edge, as imaged by
collecting the scattered ions of the FIB. The polished region (on the right) looks smoother;
a columnar pattern is visible only in the lower half of the wire.
Contrast is given by the ion
reflection coefficient, varying with the relative orientation of
the crystal axes of the gold grains.
The picture shows a small region of $9.2\,\mathrm{\mu m}\times 5.1\,\mathrm{\mu m}$ at the end of the polished section.}
\end{center}
\end{figure}

Figure~\ref{fig:micro} shows how the polished and the untreated edges of a gold conductor look like.

Experiments to study the potentials created by the sculptured wire were carried out in our atom chip BEC apparatus
described in~\cite{Sch03b}. Typically $10^{8}$ $\mathrm{^{87}Rb}$ atoms
accumulated in a MOT are transferred to a magnetic trap created by
a large wire underneath the atom chip and cooled to $5\,\mathrm{\mu K}$ by RF
evaporation. The resulting sample of $10^{6}$ atoms is then loaded
to the selected chip trap at the desired longitudinal position,
where a second stage of RF evaporation creates a pure quasi one
dimensional BEC which spans typically 0.6 mm in length. The BEC is
then kept for 300 ms in the trap with a close RF knife to achieve dynamic
equilibrium.

We image the atomic cloud by resonant absorption imaging, with
$3\,\mathrm{\mu m}$ resolution. To determine the cloud distance from
the surface $z$, we incline the imaging light with respect to the
chip mirror surface by $10\, \mathrm{mrad}$; for $z<150\,\mathrm{\mu m}$ this leads
to a double absorption image~\cite{Sch03b}, allowing a direct
measurement of $z$. For distances below the optical resolution the
value can be extracted from the known currents and fields. To
measure the atom density pictures are taken after 1.8 ms time of
flight.

If we transfer ultra cold thermal atoms to the area where the edge
is modified by the FIB we observe as dominating potential features
a trap \cite{DPie} and a barrier at the step of the
wire edge (fig.\ref{fig:Polished-area}, center).

When cooling further, the higher atomic density in the deep
potential dimple at one corner of the FIB modified area causes an
enhancement of the condensate growth~\cite{Stampe98}. Care has to
be taken that not all atoms accumulate there. Precisely
controlling the final steps of the RF cooling process we can
obtain a BEC in the dimple with atom numbers ranging from
$4\cdot10^{4}$ to $<100$. Its longitudinal trapping frequency
depends on the distance; for the set of experiments reported below
we measured values in the range
$40\,\mathrm{Hz}<\nu_{\parallel}<120\,\mathrm{Hz}$, with
transversal confinement up to $\nu_\perp=7\, \mathrm{kHz}$.

Using different parameters we can obtain an extended 1D-BEC. The
density of such a 1D-BEC is a very sensitive measure of the
potential variations at the bottom of an elongated trap and allows
to deduce the magnetic field variations along the
wire~\cite{wild05,wild06}.  We employed this measuring method to
study the potential roughness~\cite{kru05} and we use it to
compare the polished section of the wire with the unpolished
section.

Experiments were done with trapping currents ranging from $I=20\,
\mathrm{mA}$ to $I=60\, \mathrm{mA}$ at an external bias field of $0.43\, \mathrm{mT}$, leading
to atom-surface distances between $z=6.4\,\mathrm{\mu m}$ and $z=28.5\, \mathrm{\mu m}$.
Transversal confinement varied from $\nu_{\perp}=6.6\, \mathrm{kHz}$ to
$\nu_{\perp}=2.1\, \mathrm{kHz}$, longitudinal confinement was
$0.5\, \mathrm{Hz}<\nu_{\parallel}<2.0\, \mathrm{Hz}$  and trap bottom $ 86\, \mathrm{\mu T}$.
Using the modulation in the observed 1D atomic density we can
reconstruct the potential modulation
over the whole length of the 1D BEC. Over nearly the whole range
of heights, the polished section of the wire gives a smoother
potential, as characterized by the rms potential roughness. For
the lowest heights the full potential corrugation in the untouched
region cannot be measured because it can be bigger than the
chemical potential, and we exclude these data from further
analysis. Comparing the standard deviation of the potential
corrugations of polished section to the bare wire we find
${\sigma_\textrm{polished}}/{\sigma_\textrm{bare}}=0.63\pm0.1$.
The polished section gives nearly a factor two smoother potential,
indicating a contribution of wire edge roughness~\cite{Wang04} of the
same order of that due to other causes.

These measurements have to be compared to previous
observations of potential roughness.

Significant disorder in atom chip potentials has been reported at
surface distances below $\sim100 \,\mathrm{\mu m}$ when using
electroplated chip wires~\cite{For02,Lea03,Jon03,Est04}. The
production process used for such chips tends to generate columnar
structures along the edges of the conductors, with roughness around $1\,\mathrm{\mu m}$. As concluded
in~\cite{Est04} irregular current flow due to edge roughness
\cite{Wang04}, is the predominant cause for the disorder potential in wires with a width of a few $10\,\mathrm{\mu
m}$.

Atom chips built by lithographically patterned evaporated gold
wires~\cite{Gro04} on Si substrates on the contrary can show edge
roughness smaller than the scale of the
gold grains. The observed disorder potential in these chips can
be orders of magnitude smaller and is caused by the local
properties of the conductor; edge effects are negligible in many cases
~\cite{kru05}. These wires allow to create continuous 1D
condensates at distances $z$ down to a few $\mathrm{\mu m}$. The
potential roughness varies from chip to chip, and even from wire
to wire.

The specific wire used for these series of experiments shows a
disorder potential more than a factor 10 smaller than observed in
\cite{For02,Lea03,Jon03,Est04,schumm05}, but about a factor 10
larger than for the wires studied in~\cite{kru05}. Our experiments
indicate that edge roughness can give a contribution to disorder potential even for these lithographically patterned evaporated
gold wires.  In the wire studied here the contribution of the edge
is of the same order of magnitude of what comes from the local
properties of the metal.

To significantly decrease potential roughness the overall
homogeneity of the conductor, determining the current flow also in
the bulk of the wire, is the critical technological issue. We will
explore this dependence in a forthcoming experiment. To fully
exploit the possibilities given by the potential modulations
discussed below it is however clear that wires giving trapping
potentials as smooth as those seen in~\cite{kru05} should be
used.

In the following we give a few examples for other potentials which
can be created by sculpturing the wires.

\section{A double barrier potential}

\begin{figure}[t]
\begin{center}\includegraphics[scale=0.43]{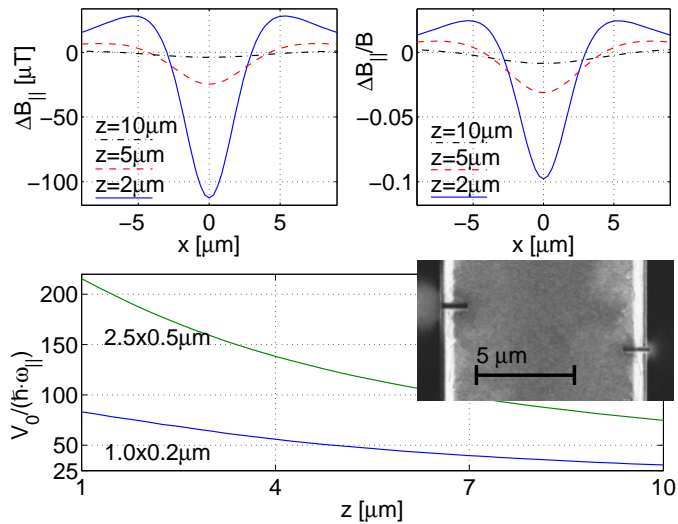}\end{center}
     \caption{(Color online) \emph{Top}: Longitudinal magnetic field modulation $\Delta B_\parallel$ in
     a 1D trap above the center of a $10\,\mathrm{\mu m}$ wide $0.25\,\mathrm{\mu m}$
     thick wire, created by two $2.5\,\mathrm{\mu m}\times0.5\,\mathrm{\mu m}$
     notches separated by $2\,\mathrm{\mu m}$, $I=25\,\mathrm{mA}$, $2\,\mathrm{\mu m}<z<10\,\mathrm{\mu m}$;
     \emph{(left)} value given in Tesla and \emph{(right)} relative to wire field B.
     \emph{Bottom}: total depth of the longitudinal potential modulation $V_0$ in
     units of the longitudinal level spacing $\hbar \omega_{\parallel}$ obtained
     by making a harmonic approximation at the potential minimum.  Data for $1\,\mathrm{\mu m}<z<10\,\mathrm{\mu m}$,
     for two different notches' dimensions. Insert: two
     notchess ( $0.9\,\mathrm{\mu m}\times0.2\,\mathrm{\mu m}$,
 $2.2\,\mathrm{\mu m}$ separation) realized (FIB)
     in a $10\,\mathrm{\mu m}\times2.5\,\mathrm{\mu m}$ wire.\label{fig:DoubleCut}}
\end{figure}

A short and deep cut can be seen as a structure generating a \textquoteleft current  dipole\textquoteright, with
components perpendicular to the normal flow. As an example two
$1\,\mathrm{\mu m}$ deep, $0.2\,\mathrm{\mu m}$ long cuts in opposite edges of the
trapping wire give rise to a double-barrier potential, with a deep
minimum in between (fig.\ref{fig:DoubleCut}).  For equal cuts, the
double-barrier is symmetric in the wire center. Asymmetries can be
introduced when moving off center. For atom transport through such
a structure resonances of the atom transmission are expected,
depending on the atom cloud density and on the number of bound
states in the minimum \cite{DPie}. In addition other non-linear effects can
become dominant~\cite{qtra}. Inverting the relative position of
the cuts changes the sign of the potential.

In correspondence with the longitudinal magnetic
field modulation $\Delta B_\parallel$, variations are generated
also in the other two field components, parallel to the chip
surface $B_y$, and  perpendicular to it $B_z$. They represent
however a negligibly small perturbation to the local field, and are
compensated by minute deviations of the trap position.  For the
cases considered here the displacement is smaller than the size of the ground state in the  transversal harmonic potential. A detailed numerical analysis shows
that the potential modulation obtained with the simple calculation, taking into account only the field $B_\parallel$ along x, and the full calculation, taking into account all three field components, differ only by a few percent.

\section{A magnetic lattice}

\begin{figure}[h]
\includegraphics[scale=0.47]{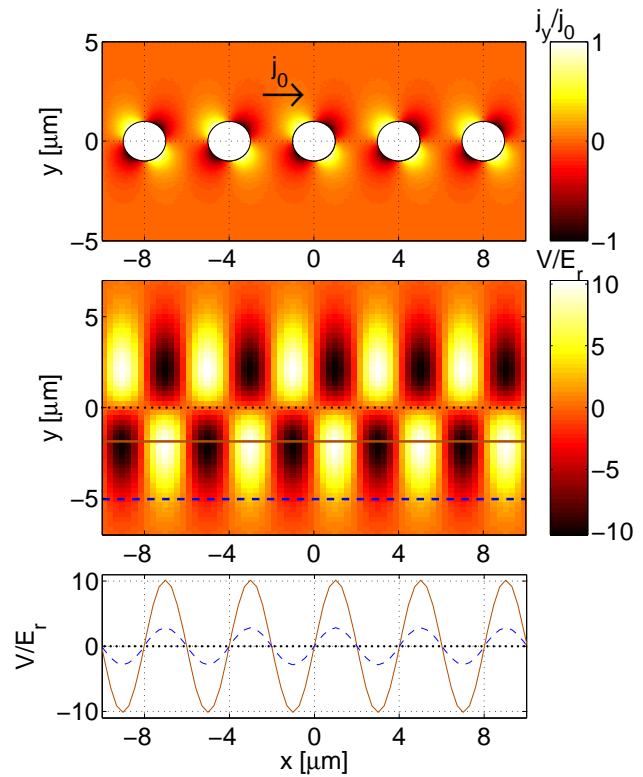}
     \caption{(Color  online) \emph{Top}: current density $j_y$ in a section of the wire where holes have been cut; $j_0$ is the current density in absence of the defects. \emph{Center}: The potential modulation $V$ at $z=6\,\mathrm{\mu m}$, in recoil energy units ($E_r={\displaystyle \hbar^2 k_a^2}/{\displaystyle  2m_{Rb}}$ with  $k_a=2\pi/a$ the lattice vector, and $a$ the lattice length). \emph{Bottom}: The potential along the lines drawn in the central image; at $y=0$ also $V=0$ (black dotted line); the maximum modulation on $V$, with a peak-to-peak depth $V_0\simeq 20\,E_r$, is obtained at $y=2\,\mathrm{\mu m}$  (orange continuous curve) while $V_0\simeq 6\,E_r$ for $y=5\,\mathrm{\mu m}$   (blue dashed curve). In all the images the lattice length is $a=4\,\mathrm{\mu m}$, $j_0=1\cdot 10^{10}\,\mathrm{A/m^2}$, the holes diameter $D=a/2$, the wire thickness $d=a/2$ and width $w=10\,\mathrm{\mu m}$.\label{fig:lattice2}}
\end{figure}

\begin{figure}[h]
\includegraphics[scale=0.43]{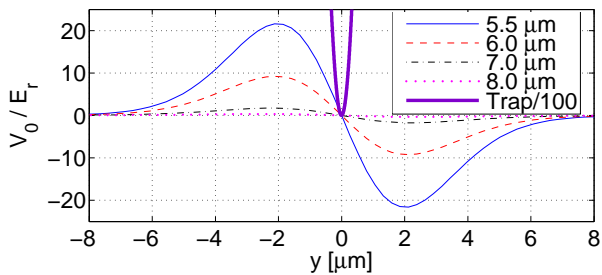}
     \caption{(Color  online) Amplitude of the peak-to-peak potential depth $V_0$ along the direction $y$ transversal to the wire; the curves, from bigger to smaller in modulus, correspond to the heights ${z=[5.5,\:6,\:7,\:8]\,\mathrm{\mu m}}$; simulations done with  ${j_{0}=1\cdot10^{10}A/m^{2}}$, $a=4\,\mathrm{\mu m}$,
     ${d=a/2}$, ${D=0.6\, a}$, ${w=10\,\mathrm{\mu m}}$. The thick violet curve shows for comparison the trapping potential for $\nu_\perp=20\,\mathrm{kHz}$, scaled down a factor 100.\label{fig:lattice1}}
\end{figure}

Extending the above idea, a regularly spaced set of cuts on the
edge, or an array of holes in the center of a plane conductor, as
illustrated in fig.~\ref{fig:lattice2}, will create a periodic
modulation of the longitudinal magnetic field, and consequently a
lattice. Compared with using a thin meandering wire, periodic
sculpturing of a wider wire has the advantage of resulting
in a much more robust structure. Moreover it allows to create both the 1D
trap and the modulation from a single setup.

Making holes in the wire center instead of cuts in its edge gives
the additional feature that the potential modulation is
antisymmetric as a function of the direction transversal to the
wire $y$.  The amplitude of the modulation and its phase depend
on the position of the 1D trap relative to the line connecting the
center of the holes. Directly above the center line the amplitude
is zero. The maximum potential modulation is found at a transverse
distance of about $y=\pm D$, where $D$ is the diameter of the hole
(see also fig.~\ref{fig:lattice1}). The modulation along the trap
axis is to a good approximation sinusoidal; the higher Fourier
components are strongly damped with height above the chip. The
potential on either side of the line of holes exhibits a $\pi$
phase shift, but is otherwise identical. In an experiment this
allows very easily to remove or invert the modulation by
a transverse translation of the 1D trap.

The actual peak-to-peak amplitude of the potential modulation $V_0$ along
$y$ is plotted  in fig.~\ref{fig:lattice1}, for various heights
$z$.  The convenient unit of energy to discuss these lattices is
$E_r = {\displaystyle \hbar^2 k_a^2}/{\displaystyle
2m_{Rb}}$, the recoil energy associated with the lattice vector
$k_a=2\pi/a$ ($a$ is the lattice period).

\begin{figure}[t]
\begin{center}\includegraphics[scale=0.43]{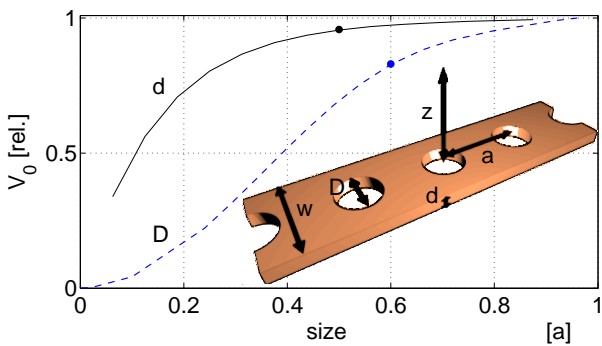}\end{center}
     \caption{(Color  online) Relative potential modulation amplitude for various
     $d$ (continuous curve) and for various $D$ (dashed curve), in units of lattice length $a$;
     the dot on the curves indicates the value used in the following simulations. Apart from the scanned parameter, the others are:  ${ j_{0}=1\cdot10^{10}A/m^{2} }$, ${ \nu_{\perp}=20\,
 \mathrm{kHz} }$, ${ a=4\,\mathrm{\mu m} }$,
      ${d=a/2}$, ${D=0.6\, a}$,
 ${ w=10\,\mathrm{\mu m} }$ ${ z=6.75\,\mathrm{\mu m} }$.  \emph{Insert}: a sketch of a wire section with the parameters used: $d$
     wire thickness, $w$ wire width, $D$ holes' diameter, $a$ lattice step, $z$
     distance atoms-chip.\label{fig:lattice0}}
\end{figure}

To maximize $V_{0}$ the diameter
$D$ of the holes should be as big as possible
(fig.\ref{fig:lattice0}, dashed curve). In the case considered
below we choose $D=0.6\, a$, a value at which they are still
nicely separated and easy to fabricate. Keeping a constant current
density, $V_{0}$ starts to saturate at a wire thickness of
$d\simeq{a}/{3}$, as expected from eq.\ref{eq:c}.
(fig.\ref{fig:lattice0} continuous curve).

To assess if one can observe the transition between a 1D super
fluid (SF) to Mott-insulator (MI) in such a magnetic lattice we
follow the work by Zwerger~\cite{Zwerg03}. For a 1D system of
bosons (chemical potential $\mu\ll\hbar\omega_{\perp}$) in a potential lattice the
crossover from the SF to the MI regime happens at $\left . \frac{U}{J}
\right |_{c}=3.84$ for an occupation of $\overline{n}=1$ atoms per lattice
site, where $J$ is the hopping amplitude, $U$ the on site interaction.
For $\overline{n}\gg1$, $\left . \frac{U}{J}\right|_{c}=2.2\,\overline{n}$.
In the case of a 1D BEC of $\;\mathrm{^{87}Rb}$ one finds in the deep lattice limit ($V_{0}\gg E_{r}$):

\begin{eqnarray}
    J &=& \frac{4}{\sqrt{\pi}}\cdot E_{r} (V_{0}/E_r)^{3/4}\exp\left(-2\sqrt{V_{0}/E_r}\right)\nonumber \\ \nonumber
    U &=& g_{1D}\int{\left|\phi(x)\right|^{4}dx}=4\pi^{3/2}\cdot \frac{a_{s}}{a}\hbar\nu_{\perp}\left(V_{0}/E_r\right)^{1/4} \\ \label{eq:e}
    \frac{U}{J} &\simeq& 3.6\cdot\nu_{\perp}a\frac{\exp\left(2\sqrt{V_{0}/E_r}\right)}{\sqrt{V_{0}/E_r}}\\ \nonumber
\end{eqnarray}
with $a_{s}$ the scattering length~\cite{kempen:02}, $\nu_{\perp}$ the trap
frequency, $a$ the lattice spacing, $\phi(x)$ the Gaussian ground
state in the local oscillator potential. Unlike~\cite{Zwerg03} the 1D nature of the trapping enters here the calculations through the 1D effective interaction strength $g_{1D}$ in eq.~\ref{eq:e}, linearly dependent on $\nu_\perp$.

\begin{figure}[t]
\begin{center}\includegraphics[scale=0.43]{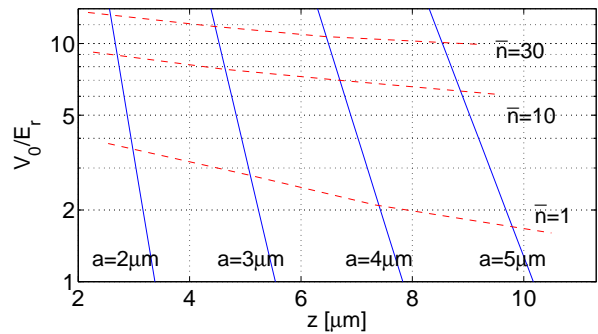}\end{center}
     \caption{(Color  online) Maximum potential modulation in lattice recoil
     energy units at ${2\,\mathrm{\mu m}<z<10\,\mathrm{\mu m}}$; continuous blue curves correspond to
      (left to right) ${a=[2,\:3,\:4,\:5]\,\mathrm{\mu m}}$; dashed red lines:
     critical value of $V_{0}$ for the Mott insulator transition, for (bottom to top)
     ${\bar{n}=[1,\:10,\:30]}$ atoms per lattice site and keeping a constant ${\nu_{\perp}=20\, \mathrm{kHz}}$ by modulating the trap bottom. Simulations done
     at current density ${j_{0}=1\cdot10^{10}\mathrm{A/m^{2}}}$,
     ${d=a/2}$, ${D=0.6 a}$, ${w=10\mathrm{\mu m}}$~.\label{fig:lattice}}
\end{figure}

For $\nu_{\perp}=20\, \mathrm{kHz}$, $d=a/2$, $D=0.6\, a$ and $2\,\mathrm{\mu m}<a<5\,\mathrm{\mu m}$
the SF-MI crossover can be achieved at $z\simeq1.5\,a$
(fig.\ref{fig:lattice}).  The characteristic timescale for an
experiment is given by the single particle tunnelling time; for
$\overline{n}=1$ it lies in the range of a few ms, about 2
orders of magnitude shorter than the expected lifetime at such $z$
\cite{Zhang05,Henkel,Henkel1,Henkel2,Henkel3,Rekdal04,Scheel05}. The limiting factor in an
experiment would be detecting the low 1D atoms' density
\cite{deteff}.

Increasing $\overline{n}$ up to 30 requires $\left . \frac{U}{J}\right |_{c}=66$
and the transition occurs around $V_{0}\simeq12\,E_{r}$
(fig.\ref{fig:lattice}, upper red dashed curve), with a single particle tunnelling
time in the order of tens of ms, still about one order of
magnitude shorter than the estimated lifetime and at a density
experimentally observable.

The required value of $V_{0}$ at the transition can be lowered by
moving deeper into the 1D regime, for example
increasing $\nu_{\perp}$; keeping a minimum trap bottom $B_x$ of $50\,\mathrm{\mu T}$ we can reach a confinement of $\nu_\perp=80\,\mathrm{kHz}$. In this case the values for $U$ and $J$
used above (deep lattice) can be no longer valid. In the limit of a
weak lattice, raising the transverse confinement we move into the
Tonks-Girardeau (TG) regime, increasing the ratio between
interaction and kinetic energy per particle: $\gamma=mg/(\hbar^2
n_{1D})$, with $g=2\hbar a_s\cdot 2\pi\nu_\perp$ and $n_{1D}=\bar{n}/a$
\cite{Zwerg03,BlochTG}. In the analytically solvable case of one
atom per lattice site, a finite excitation gap $\Delta$ is
observable before the deep TG regime~\cite{Zwerg03}. Keeping a
minimum trap bottom at $70\, \mathrm{\mu T}$ to avoid spin flips, and
$j_0=1\cdot 10^{10}\,\mathrm{A/m^2}$, we can achieve $\gamma > 10$ for $z\leq
2\,a$; $\Delta$ is in this case of the same order of magnitude of $V_0$.

The above estimates are rather conservative. Our atom chips
\cite{Gro04} can support at least 10 times larger current
densities, allowing either thinner wires resulting in a 10 times
longer spin flip lifetime or much tighter confinement so that reaching
$\gamma\geq10^4$ should be possible.

Increasing the $\gamma$ factor while having at the same time $\bar{n}>1$
 can lead to novel phenomena
\cite{alon:05}: atoms in a MI phase
 can undergo a local TG-like transition in each lattice site and localize in spatially separated distributions.

\section{Conclusion}
We have used the characteristics of small changes in
the current flow to enhance the flexibility in the design and
implementation of tailored potentials for complex matter wave
manipulation on atom chips.  In a pilot experiment we have
demonstrated engineered atom chip potentials and shown that the
potential roughness can be improved by polishing the wire edge
using FIB milling.  These sculptured wires hold promise for many
microscopic atom optical potentials on atom chips, like double
wells, double barriers and magnetic lattices. An intriguing
possibility is to create a potential with designed disorder or
defects, combined with single site addressability and control on the atom chip.

This work was supported by the European Union, contract numbers
IST-2001-38863 (ACQP), HPRN-CT-2002-00304 (FASTNet), and
RITA-CT-2003-506095 (LSF), the Deutsche Forschungsgemeinschaft,
contract number SCHM 1599/2-2 and the German Federal Ministry of
Education and Research (BMBF) through the German-Israel Project
DIP-F 2.2.


\begin{thebibliography}{30}
\bibitem{PotSens}
    Ultra cold atoms are very sensitive to field variations. For $\mathrm{^{87}Rb}$
    in the F=2, $m_{f}=2$ state $0.1\,\mathrm{\mu T}$ corresponds to 67 nK. At a
    bias field of $1\,\mathrm{mT}$ this corresponds to a change in the current
    flow direction of $10^{-4}$ rad.
\end{thebibliography}
\end{document}